\documentclass[twocolumn,epsfig,showpacs,superscriptaddress,preprintnumbers,amsmath,amssymb]{revtex4}

\usepackage{graphicx}
\usepackage{dcolumn}
\usepackage{bm}
\usepackage{epsfig}

\def\prb{Phys.\ Rev.\ B}

\def\prl{Phys. Rev.\ Lett.}

\def\be{\begin{equation}}
\def\ee{\end{equation}}
\def\ba{\begin{eqnarray}}
\def\ea{\end{eqnarray}}

\def\LSCO{La$_{2-x}$Sr$_x$CuO$_4$}
\def\LCO{La$_2$CuO$_4$}

\def\YBCO{YBa$_2$Cu$_3$O$_{7-\delta}$}

\def\BSCCO{Bi$_2$Sr$_2$CaCu$_2$O$_{8+\delta}$}
\def\C60{A$_x$C$_{60}$}
\def\LNSCO{La$_{1.6-x}$Nd$_{0.4}$Sr$_x$CuO$_{4}$}

\def\HgCu3{HgCa$_2$Cu$_3$O$_{8+y}$}
\def\HgCu4{HgBa$_2$Ca$_3$Cu$_4$O$_{10+y}$}
\def\TlCu3{Tl$_2$Ba$_2$Ca$_2$Cu$_3$O$_{10+y}$}
\def\TlCu4{Tl$_2$Ba$_2$Ca$_3$Cu$_4$O$_{12+y}$}

\def\BiCu3{Bi$_2$Sr$_2$Ca$_{2}$Cu$_3$O$_y$}

\begin{document}

\title{Competing Order in the Mixed State of High Temperature Superconductors}

\author{Steven A.~Kivelson}
\affiliation{Department of Physics,
University of California at Los Angeles,
405 Hilgard Ave.\ , Los Angeles, CA 90095}
\author{Dung-Hai Lee}
\affiliation{Department of Physics, University of California, Berkeley, CA 94720}
\author{Eduardo Fradkin}
\affiliation{Department of Physics, University of Illinois,
1110 W.\ Green St.\ , Urbana IL 61801}
\author{Vadim Oganesyan}
\affiliation{Department of Physics, Princeton University,
Princeton, NJ 08544}

\date{\today}

\begin{abstract}
\

We examine the low temperature behavior of the mixed state of a layered
superconductor in the vicinity of a quantum critical point separating a pure
superconducting phase from a
phase in which a competing order coexists with superconductivity.
At zero temperature, we find that there
is an avoided critical point in the sense that the phase boundary in the limit
$B\to 0$ does not connect to
the $B=0$ critical point.  Consequently, there exists a quasi-1D regime
of the phase diagram, in which the
competing order is largely confined to 1D ``halos'' about each vortex core,
and in which interactions
between neighboring vortices, although relevant at low temperature, are relatively weak.
\
\end{abstract}

\maketitle
\narrowtext

Whereas in many well understood metallic compounds  over a
broad range of compositions and temperatures, the only two phases
encountered are the normal (Fermi liquid) and superconducting phases, in
the cuprate high temperature superconductors, and other
highly correlated electronic systems, there are many ordered phases which
appear to compete and sometimes coexist.  In addition to the uniform
d-wave superconducting state, compelling evidence exists of ordered
antiferromangetic (N\'eel), unidirectional charge density wave
(``charge-stripe''), unidirectional, colinear, incommensurate
spin-density wave (``spin-stripe'') phases and those with coexisting
superconducting and stripe order~\cite{PNAS}.  Preliminary evidence also
exists of possible $d$-density-wave or staggered flux order~\cite{dai,sudip}, electron
nematic~\cite{ando}, $d+id$ or $d+is$ superconducting order~\cite{laura,deutcher},
and various other
phases in which more than one of these orders coexist.

\begin{figure}[bht]
\begin{center}
\leavevmode
\vspace{.2cm}
\noindent
\hspace{0.3 in}
\centerline{\epsfxsize=3in \epsffile{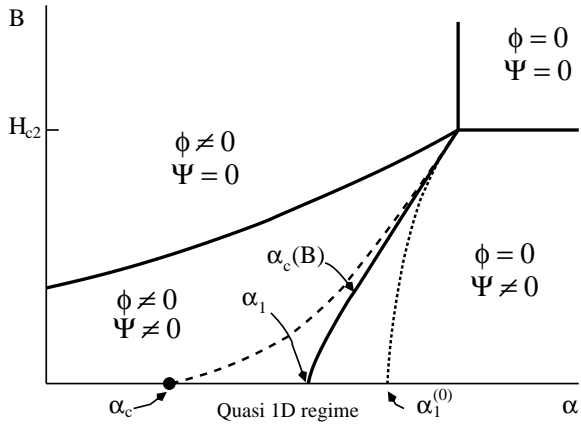}}
\end{center}
\caption
{Schematic zero temperature phase diagram of a layered superconductor as a function 
of a control parameter $\alpha$ (described in the text) and the magnetic induction
$B$, {\it i.e.} the actual magnetic field which penetrates the system.
 Here $\Psi$ and $\phi$ denote, respectively, the expectation values
of the superconducting and competing order parameters and it is assumed that for $B=0$
there is a
continuous transition at $\alpha=\alpha_c$ between a pure superconducting phase and a
phase with a coexisting
$\Psi$ and $\phi$ order.
The solid lines denote phase boundaries and the other lines are crossovers (all are
described in the text).
In particular, $\alpha_c(B)$ represents the boundary between the pure superconducting
and the phase with coexisting
$\phi$ and superconducting order, and $\alpha_1\equiv \lim_{B\to 0} \alpha_c(B)$ marks
the point at which the $\phi$
``halo'' about an isolated vortex-line undergoes a transition from a quantum disordered
state (for $\alpha > \alpha_1$) to
an ordered or quasi-ordered state for $\alpha < \alpha_1$.  ($\alpha_1^{(0)}$ is the
mean-field value of $\alpha_1$.)}
\label{phasediagram}
\end{figure}

One rather direct way to look for competing order parameters in a
system which is globally superconducting was recently proposed by Zhang and
coworkers~\cite{zhang,arovas}
and others~\cite{palee,dhlee,demler}.  The idea
is that if the superconducting order
is only slightly favored over a competing order, then where the superconducting
order is suppressed in the core of a vortex, the competing order will be manifest. Indeed,
recent neutron scattering experiments of Lake {\it et.\ al.\/}~\cite{lake} on {\LSCO} have
revealed a strong enhancement of ``spin-stripe'' order at low energies
produced by modest magnetic fields.  Similar results have been obtained in {\LCO} by
Birgeneau {\it et.\ al.\/}\cite{birgeneau}.  Moreover, in
scanning tunnelling microscopy (STM) studies by J.\ Hoffman {\it et.\ al.\/} \cite{davis}
of near optimally doped {\BSCCO} in a 7T magnetic field,
large induced ``halos'' about each vortex core have been imaged where the density of states
is modulated with a spatial period
(4$a$) equal to that expected\cite{tranquada} for ``charge-stripe'' order.
Additional evidence that there is a substantial degree of local charge stripe order in
{\BSCCO}
comes from the similar patterns of density of states modulation observed in the
zero-field STM studies of C.\ Howald {\it et.\ al.\/}~\cite{kapitulnik} on
the same material. (This interpretation is, however, being challenged in a forthcoming
paper by Hoffman {\it et al}\cite{newdavis}.)

The notion of a competing order developing an expectation value in a vortex core is
basically a mean-field notion.
However, in all the examples mentioned above, the competing order is
associated with spontaneously broken symmetry, so that fluctuation effects may
fundamentally alter the physics.
Specifically, in a planar system, the vortex core is a finite size system, and so cannot
support a spontaneously broken
symmetry, while in three dimensional superconductors the vortex core is a one dimensional
system, so cannot exhibit any
symmetry breaking except at $T=0$, and there only for discrete symmetries.

\section{Phase Diagram}

Our principal results are summarized in the schematic zero temperature phase diagram
(Fig.\ \ref{phasediagram})
for a layered system in which superconducting order, with an order parameter denoted
by $\Psi$, competes with
another type of order, whose order parameter is denoted by
$\phi$.   The two axes represent the magnetic induction, $B$, and a control parameter,
``$\alpha$,'' such as pressure or
doping concentration, with the convention that increasing $\alpha$ disfavors
$\phi$ order.  It is assumed that at $B=0$ there exists a continuous quantum phase
transition at
$\alpha=\alpha_c$ (the heavy circle) separating a pure superconducting phase from a
phase with coexisting
$\phi$ and superconducting order.

While the considerations here are rather general, it is useful to have specific
realizations in
mind.  For us, the most important case is that in which $\phi$
represents~\cite{nature} stripe
orientational (``nematic'') order. In the absence of crystal field effects,
the orientation of the stripes is arbitrary and can be parametrized by an angle
$0 \leq \theta <\pi$. Hence, in this case $\phi$ would then be a {\sl director},
a headless vector,
which in two dimensions can be represented by a complex scalar field defined so
that $\phi=|\phi| \; e^{i2\theta}$
corresponds to stripes lying along a preferred direction
$\hat e_{\theta} = \hat x \cos \theta +\hat y
\sin \theta$. (Note that $\theta=0$ and $\theta=\pi$ are physically equivalent.)
In this limit, $\phi$ has a continuous XY symmetry.
However, in a  crystal with appropriate point-group symmetry~\cite{comment-2}
there are only two preferred stripe orientations. In this case, $\phi$ reduces to a
real scalar field reflecting the Ising character of
the symmetry breaking.

If, on the other hand the relevant stripe order is magnetic (incommensurate SDW) order, in addition to
orientational symmetry the relevant broken symmetry is spin rotational invariance, so for each stripe orientation,
$\phi$ is a three component real vector field corresponding to Heisenberg symmetry. We shall see that this case is
slightly different than the lower symmetry situations.  Stripe states can also spontaneously break translational
symmetry, but since a vortex core explicitly breaks this symmetry in any case, issues of spatial symmetry breaking
are more subtle, and will be discussed elsewhere~\cite{later}.

In the phase diagram shown in Fig.\ \ref{phasediagram}, the solid lines represent actual
phase boundaries, and the broken lines are crossovers.  The most striking feature of this
phase diagram is the presence\cite{avoid} of an ``avoided critical point,''
{\it i.\ e.\/} the phase boundary $\alpha_c(B)$ has a discontinuity at $B=0$:
\begin{equation}
\lim_{B\to 0} \alpha_c(B) \ne \alpha_c
\end{equation}
This discontinuity in the phase boundary is a consequence of the fact that the magnetic
field is a singular perturbation.

The assumed competition (mutual suppression) between
superconducting and
$\phi$ order (the latter assumed to be essentially uncoupled to $B$) is seen in the fact
that the critical line,
$\alpha=\alpha_c(B)$ separating the pure and coexistence phases is an increasing function
of $B$ up to the point
at which superconductivity is completely suppressed by a magnetic field in excess of
$B=H_{c2}$.
The dashed line marks a crossover from quasi one dimensional to fully three dimensional
order (where $\phi$ is
more or less uniform in space). It essentially coincides with the phase boundary derived
by Demler {\it et.\
al.\/}~\cite{demler} for the case in which interactions between layers are
neglected\cite{2dcase}. The
dotted line represents a crossover associated with the mean-field phase boundary;
just to the left of
this line, there is a ``halo'' of the competing order surrounding each vortex core,
but each halo fluctuates
essentially independently, and there is no true
$\phi$ order.

The arguments that lead
to this phase diagram constitute the bulk of the present paper.
The reader should be warned that
 there are some subtleties (which we will discuss below) associated with
one or another specific type of competing order that can affect the shape,
and even the topology of the phase diagram, as does even weak quenched disorder.
For simplicity
of discussion, for most of the paper we will assume the extreme type II limit,
in which the London penetration depth, $\lambda=\infty$, although this assumption is
not necessary.  Some of the experimental consequences  of this phase diagram are
discussed in the final section of this paper.

\section{The basic physics}

In this section, we sketch the basic physics that leads to the phase diagram in
Fig. \ref{phasediagram}.
All actual derivations are deferred to later sections.  For simplicity, let us first
consider the case in
which the broken symmetry associated with the competing order is Ising-like,
{\it i.\ e.\/} $\phi$ is a real
scalar, and there is a symmetry under
$\phi\to -\phi$.

\subsection{The Ising Case}

We start our discussion by considering the structure of a single, isolated vortex in the
uniform
superconducting state~\cite{comment-3}, {\it i.\ e.\/}
$\alpha> \alpha_c$.  If $\alpha$ is large, the structure of the vortex is unaffected by
the proximity of the
$\phi$ ordered phase, but if $\alpha$ is sufficiently close to
$\alpha_c$, then at mean-field level, the suppression of
$\Psi$ in a vortex core region of size
$\xi_0$  (the superconducting coherence length) will result in a halo with radius equal
to the critical
correlation length,
$\xi_{\phi}
\gg
\xi_0$, in which $\phi\ne 0$.  An important aspect of the structure of the vortex, first
emphasized in this context by Demler {\it et.\ al.\/}~\cite{demler}, is that the magnitude
 of the superconducting order
parameter does not return to its bulk value exponentially, but rather, for a vortex at the
origin and for $r\gg\xi_0$, has a power-law
form
\begin{equation}
|\Psi(\vec r)|^2 \sim \Psi_0^2\; [ 1 - (\xi_0/r)^2 +\ldots] .
\end{equation}
This  $1/r^2$ fall-off is a necessary consequence of the slow decay of the
superfluid flow around an isolated vortex, and results in a somewhat larger halo of
$\phi$ order
around a vortex than would otherwise occur. In any case,
$\xi_{\phi}\sim |\alpha-\alpha_c|^{-\nu}$ diverges with a critical
exponent $\nu$ as $\alpha\to \alpha_c$. Since $\nu$ is the quantum critical exponent
of a system in $d=3$ spatial dimensions, it
presumably takes its mean-field value, $\nu=1/2$ (up to logarithmic corrections-to-scaling).

Within a single superconducting plane the halo is a finite size system, and so cannot
actually support a broken symmetry state, {\it i.\ e.\/} quantum fluctuations will cause
the system to tunnel
between the positive and negative $\phi$ states with a matrix element $h\equiv h(\alpha)$.
Thus, in the absence of inter-plane couplings, there
can be no true $\phi$ order established until $B$ is large enough, {\it i.\ e.\/} the vortex
density is large enough, that there is significant
coupling between neighboring vortices.   Because the size
of the halo increases with increasing $\xi_{\phi}$, we expect that $h$ vanishes
in the limit
$\xi_{\phi}/\xi_0\to \infty$. We derive an explicit expression for this, below
in Eq.\ \ref{eq:h},  from
the Landau-Ginzburg theory of
competing orders.
However, even if the microscopic coupling between the order parameter $\phi$ on
neighboring planes, $J_0$, is weak, the large size of the halo implies
a large effective coupling, $J$, between the halos on neighboring planes, with $J \sim J_0
{\overline{\phi^2}}\xi_{\phi}^2$, where $\overline{{\phi^2}}$ is the mean squared value
of $\phi$ in the vortex halo.  {From} the Landau-Ginzburg theory, this coupling is
seen to diverge as $\xi_{\phi}/\xi_{0}\to \infty$, although
only logarithmically, $J \propto J_0 \log(\xi_\phi/\xi_0)$, as shown in Eqs.\ (\ref{moments})
and (\ref{J}).

Thus, along an isolated vortex, the low energy fluctuations
are equivalent to an effective transverse-field Ising model with Ising coupling $J$
and transverse field $h$.
This model has an ordered ground-state provided $J/h > 1$.  Since as $\alpha$ approaches
$\alpha_c$, the limit $J/h\to \infty$ is realized,  it follows that for $\alpha$ near
$\alpha_c$, an isolated vortex line has an ordered
ground-state!  Moreover, in this limit, no matter how small
$B$, the inter-vortex coupling can never be ignored, and indeed leads to a finite
transition temperature.  We can
estimate this transition temperature using standard methods~\cite{scalapinoferrel}
of quasi-one dimensional
systems:  We estimate the inter-vortex coupling to be
$J_{inter}\sim
\exp(-r/\xi_{\phi})\sim
\exp(-A/\sqrt{B\xi_{\phi}^2})$, where $r$ is the spacing between vortices and $A$ is
a number of order 1.
At low temperatures, we estimate the susceptibility of an isolated vortex line to be
that of the 1D Ising chain, $\chi_{\phi}(T)\sim
\exp(A^{\prime}J/T)$.  Finally, we estimate
$T_c$ according to
$\chi_{\phi}(T_c)J_{inter}=1$.  This leads to the estimate
\begin{eqnarray}
T_c(B) &&\sim \xi_{\phi}J\sqrt{B} \\
\label{Tc-ising}
&&\sim B^{1/2}|\alpha-\alpha_c|^{-1/2}\log[\alpha_c/|\alpha-\alpha_c|]
\nonumber
\end{eqnarray}
where in the second expression we have adopted the Landau-Ginzburg estimate of $J$.

Reversing the present logic, it is clear that with increasing $\alpha$, we will
eventually encounter
a condition in which $h > J$. Here, the ground-state of the isolated vortex line
is quantum disordered, and
hence even at $T=0$, there is no $\phi$ ordering until a critical field strength,
$B_c(\alpha)$, is exceeded,
such that the interactions between neighboring vortices is strong enough to induce
ordering.
The critical value
$\alpha_c(B) \to \alpha_1$ in the limit $B\to 0$ marks the quantum critical point of the
isolated vortex, where
$J(\alpha_1)/h(\alpha_1)=1$.  (For small $|\alpha-\alpha_1|$, the phase boundary is
non-analytic\cite{critical}
 - a subtlety which we have neglected in sketching Fig.\ \ref{phasediagram})

The remaining phase boundaries that occur at larger values of $B$ are determined by the
obvious and
conventional physics of competing orders.   We now discuss the physics of the two
crossover lines shown in
Fig.\ \ref{phasediagram}.

Where the $\phi$ halos about the vortices start to overlap strongly, there is a crossover
from a
quasi-one dimensional regime in which the magnitude of $\phi$ is substantially
inhomogeneous, to a regime
where the variations of $\phi$ are relatively small.  (Indicated by the dashed line in
Fig.\ \ref{phasediagram}.)  At
first guess might be that this crossover occurs when the spacing between vortices is of
order $\xi_{\phi}$
or in other words when
$B^* \sim |\alpha-\alpha_c|^{2\nu}=|\alpha-\alpha_c|$ (up to logarithmic corrections to
scaling). In fact, as shown by Demler {\it et.\ al.\/}~\cite{demler}, this crossover occurs
at
somewhat smaller
$B$ due to the slow recovery of $\Psi$ away from the vortex core.
If we consider the case in which $\phi$ is homogeneous, then it is as
if there were a magnetic field dependent reduction of the effective
$\alpha_{\rm eff} = \alpha-\tilde\gamma
|B|\log|H_{c2}/B|$ where, as we will see in the next section, $\bar \gamma$ is proportional
to the coupling
strength between the two order parameters.  Consequently
$B^*\sim |\alpha-\alpha_c|/\log|\alpha_c/(\alpha-\alpha_c)|$.  (Precisely at
$\alpha=\alpha_c$, the same line of reasoning leads to the conclusion that
$T_c\sim |B\log(H_{c2}/B)|^{\nu z}$, where
$z$ is the dynamical scaling exponent, and hence depends on the dynamics of $\phi$.)

For large enough $\alpha>\alpha_1^{(0)}$, the competing order is sufficiently disfavored that, even at mean field
level, no halo of non-zero $\phi$ is induced about the vortex core.  Thus, for small $B$ and $\alpha_c(B) < \alpha
< \alpha_1^{(0)}$, there are substantial local $\phi$ correlations along each vortex, but the quantum fluctuations are
sufficiently large that the coupling between vortices can be neglected, and no true long-range $\phi$ order
develops.  For $\alpha>\alpha_1^{(0)}$, there are no substantial $\phi$ fluctuations induced by the presence of
vortices.  (The value of $\alpha=\alpha_1^{(0)}$ at which a non-zero value of $\phi$ appears at mean-field level
is computed from Landau-Ginzburg theory in Eq. (\ref{alpha1}); roughly it is the point at which $\xi_{\phi}\sim
\xi_0$.)

We note, in passing, that even in 2D the Ising case is quite different
than than Heisenberg case considered by Demler {\it et al}\cite{demler}.
In particular, as discussed in footnote \cite{2dcase}, in the Ising case the
transition to the coexistence phase occurs when the spacing between vortices is parametrically
large compared to
$\xi_{\phi}$.  Consequently, although $\lim_{B\to 0} \alpha_c(B)=\alpha_c$, for $\alpha$ close to
$\alpha_c(B)$ and $B$ small, $\phi$ is not spatially uniform but rather is strongly peaked in halos about individual
vortex cores.  

\subsection{XY and Heisenberg cases}

If $\phi$ is not an Ising variable, but has higher symmetry, the above considerations
are somewhat
modified.  In the case of an $XY$ variable, the physics of the isolated vortex is
equivalent to the
well-known physics of the 1D quantum rotor model.  Again, there is a single site
({\it i.e.} intra-plane)
term, which can be characterized by the energy gap, $h$, between the ground-state
and first excited
state of a single rotor;  manifestly,
$h$ is proportional to the inverse moment of inertia of the rotor
and large $h$ favors a quantum disordered state.  Ordering, however, is again promoted
by an ``exchange"
interaction,
$J$, between neighboring ``sites,'' {\it i.\ e.\/} neighboring planes.
As in the Ising case, we estimate the dependence
  of $J$ on $\alpha$ from the Landau-Ginzburg treatment, below, and with the same
result.  For a continuous symmetry, $h$ does not
involve tunneling, and so its dependence on $\alpha$ is much weaker than in the
Ising case;  indeed, we will
see from the Landau-Ginzburg treatment that $h$ has the inverse dependence as
$J$, {\it i.\ e.\/}
$h(\alpha) \propto 1/J(\alpha)$.

There is still a quantum disordered phase
possible if
$h/J$ is sufficiently large.  However, at smaller
$h/J$ the ordered phase of the Ising chain is replaced by a conformally invariant
(power-law) phase in the
XY chain.  The susceptibility in this phase is a power-law in temperature,
$\chi_{\phi}(T) \sim
J^{-1}(J/T)^{K}$, where $K(J/h)$ is an increasing function of $J/h$.
So long as $K>0$, this susceptibility still diverges as $T\to 0$,
so the topology of the phase
diagram is similar to that
in the Ising case.  However, now $\alpha_1$
is
determined implicitly from the relation
$K(\alpha_1)=0$.  For $\alpha_1>\alpha>\alpha_c$, where $K(\alpha)>0$,
we can use the same inter-vortex mean-field theory to estimate
the ordering temperature: 
\begin{equation}
T_c \sim J \exp[-A/(K \xi_{\phi}\sqrt{B})],
\end{equation}
{\it i.\ e.\/} $T_c$ rapidly becomes immeasurably small at small $B$.

The classical Heisenberg ferromagnet has an ordered ground state, even in one dimension.
However, the $O(3)$ quantum rotor model, which represents
the low energy theory of the quantum Heisenberg antiferromagnets, does not.
In the absence of Berry-phase terms,
the one-dimensional $O(3)$ chain  possesses~\cite{haldane} a Haldane gap, and
hence a non-divergent zero temperature
susceptibility.  This means that the topology of the phase diagram is different for the
Heisenberg case, and that there is no avoided critical point.  However, since the
Haldane gap vanishes
exponentially for large $J/h$,
$E_{Haldane}\sim
\exp[-A^{\prime\prime}J/h]$, it follows from simple scaling arguments  that
$\chi_{\phi}(T=0)\sim \exp[2A^{\prime\prime}J/h]$.  Thus, following a line of argument similar to the
one which in the Ising case led to Eq.\ (\ref{Tc-ising}), inter-vortex mean
field theory leads to an estimate for
the critical $B$,
\begin{equation}
B_c(\alpha)\sim [\xi_{\phi} J/h]^{-2}\sim
{|(\alpha/\alpha_c)-1|\over \log^{2}|(\alpha/\alpha_c)-1|},
\end{equation}
where in the second expression we have adopted the Landau-Ginzburg estimates of $h$ and $J$.

It is also possible to imagine that a single vortex corresponds to a half-integer spin
chain, in which case
there is a Berry's phase and consequently no Haldane gap.  In this case, the situation
is essentially equivalent
to the XY case, with the susceptibility exponent $K=1$ (up to logarithmic corrections).
However, it seems to us that
since the effective Heisenberg model in the present case is not sharply defined on the
lattice scale,
fluctuation effects are likely to smear out the subtle interference phenomena responsible
for the special
behavior of half-integer spins, leaving us with the physics of the rotor described above.

\subsection{Further subtleties}

There are still other subtleties to worry about.  In this entire discussion we have
assumed that the vortex texture of $\Psi(\vec r)$ does not lift the symmetries which are
spontaneously
broken by the ordering of $\phi$.  However,  where one of those broken symmetries is
translation
invariance, the presence of a vortex core is an explicit symmetry breaking field.
If this effect is
significant (as it may well be in the case of stripe order), it greatly complicates the
analysis.  A related  issue is that we
have assumed that there is no frustration of  global $\phi$ order which arises from
the form of the vortex
lattice and the nature of the coupling between neighboring halos.  We believe that,
in the absence of the
just mentioned symmetry breaking terms, this assumption is reasonable, but it is
manifestly unreasonable in
their presence.  Finally, especially in the regime where the physics is quasi one
dimensional, all results
are likely to be extremely sensitive to even tiny amounts of quenched disorder.
We have not fully
explored the implications of any of these further problems.

\section{Landau-Ginzburg Theory}

While  most features of the problem are largely determined by considerations
of order parameter symmetry, it is pedagogically useful to make them explicit by
considering the Landau-Ginzburg
treatment of two competing order parameters.  We are interested in ground-state
properties, so
we must ultimately analyze a D+1 dimensional quantum action, where D=3 is the spatial
dimension.  However, 
we will be analyzing this action semiclassically, in the sense that we will first
consider time-independent field
configurations which minimize the action, and then analyze quantum fluctuations
about this classical ground-state.
Thus, we start by considering only the
classical (static) Landau-Ginzburg  free
energy density functional in a single plane of a layered system
\begin{equation}
{\cal F}[\Psi,\phi] = {\cal F}_{SC}[\Psi]+{\cal F}_{\phi}[{\bf \phi}] +
\frac{\gamma}{2} \; |\Psi|^2 | \; {\bf \phi}|^2
+
\ldots
\label{calF}
\end{equation}
where $\Psi$, a complex scalar field, is the superconducting order parameter and
$\phi$, which
may have multiple components,  represents the competing order parameter.
In the present paper, we
will only focus on ``competing'' orders in the sense that we will always assume
that $\gamma > 0$.  Indeed,
we assume that $\gamma$ is not small - if $\gamma$ is small it means that the two order
parameters hardly interact, as happens, for instance, in
conventional superconductors
when $\phi$ and $\Psi$  originate from different pieces of the Fermi surface.
The free energy must be
invariant under
 a global U(1) transformation, $\Psi(\vec r)\rightarrow e^{i\alpha_0}\Psi(\vec r)$
 due to gauge
invariance, and, depending on the nature of the competing order, under an
additional set of global
transformations,
${\bf \phi}(\vec r)\rightarrow {\bf g} {\bf \phi}(\vec r)$ where ${\bf g}$ are
elements of an
appropriate coset space.  For instance, if ${\bf \phi}$ corresponds to N{\`e}el
order, then ${\bf g}
\in SU(2)/U(1)$.  In Eq. \ref{calF}, the superconducting contribution to ${\cal F}$
is of the usual form,
\begin{equation}
{\cal F}_{SC}= \frac {\kappa_{0}} 2 |(\frac{\vec\nabla}{i} - \frac{2e}{c}\vec A) \Psi|^2 -
\frac
{\Psi_0^2} 2 |\Psi|^2 +\frac{1}{4} |\Psi|^4 + \ldots
\label{calFSC}
\end{equation}
where $\vec A$ is the vector potential, $\xi_0
= \sqrt{\kappa_{0}/2\Psi_0^2}$
is the coherence
length, and here, and elsewhere, $\ldots$ refers to higher order terms in powers of the order
parameters.  The competing order is governed by 
\begin{equation}
{\cal F}_{\phi}= \frac {\kappa_{\phi}} 2 |\vec\nabla {\bf \phi}|^2 + \frac
{\alpha} 2 |{\bf \phi}|^2 +\frac{1}{4} |{\bf \phi}|^4
+\ldots.
\label{calFphi}
\end{equation}

The mean-field solution is obtained by solving the Landau-Ginzburg (LG)
equations, $\delta {\cal F}/\delta\Psi|_{\bar \Psi}=
\delta {\cal F}/\delta\phi|_{\bar \phi}=0$.  For the most
part, we will focus on states deep in the superconducting phase, where we can
treat $\bar\Psi(\vec r)$ as a
given function, leaving us with the task of computing $\bar \phi$.
By symmetry, $\bar\phi=0$ is always a
solution;  where a non-trivial solution exists, it leads to a condensation energy:
\begin{equation}
E_{cond} =\int d\vec r[{\cal F}(\bar\Psi,\bar \phi)-{\cal F}(\bar\Psi,0)] =-\int d\vec r
\frac {|\bar\phi|^4} 4.
\label{condensation}
\end{equation}

In the spatially uniform case, $\bar \Psi(\vec r) = \Psi_0$, there is a non-trivial
solution, $\bar\phi(\vec r)=\sqrt{(\alpha_c-\alpha)/(1-\gamma^2)}$ for
$\alpha < \alpha_c = -\gamma\Psi_0^2$, while $\bar
\phi=0$ for $\alpha > \alpha_c$.  In the $\phi$ disordered phase, it is
easy to see that any inhomogeneous
solution of the LG equations will decay exponentially with a correlation length
$\xi_{\phi}=\sqrt{\kappa_{\phi}/(\alpha-\alpha_c)}$.

In the case of a single vortex at the origin, we look for a solution of the form
$\bar\Psi(\vec r)=e^{i\varphi}|\bar\Psi( r)|$ where $\varphi$ is the azimuthal angle.
While the exact form of $|\bar\Psi( r)|$ is
somewhat complicated, at large $r$ it is easily seen to be
$|\bar\Psi(r)|=\Psi_0[1-(1/2)(\xi_0/r)^2 +{\cal O}((\xi_0/r)^4)]$.
(Of course, if
magnetic screening is taken into account, $|\bar\Psi(r)|$ approaches 
$\Psi_0$ exponentially for $r \gg \lambda$, with the London penetration length;
in the high temperature
superconductors, $\lambda$ is very large, so it is reasonable to approximate
it as infinite.)
comment at various stages on the effect of finite $\lambda$ on our results.)
Since none of our results depend
critically on the short distance ($r<\xi_0$) behavior of this solution,
we will adopt the approximation
\begin{equation}
|\bar\Psi(r)|^2= \Psi_0^2[1-(\xi_0/r)^2] \ \ {\rm for} \ \ r > \xi_0
\end{equation}
and $|\bar\Psi(r)|^2=0$ for $r < \xi_0$.  Thus, the LG equation for an isolated vortex is
\begin{equation}
\left[-\frac {\partial^2} {\partial x^2} -\frac 1 x \frac {\partial} {\partial x}
+\left(\frac{\xi_0} {\xi_{\phi}}\right)^2 -\frac {\tilde \gamma} {x^2} +
\phi^2(x)\right]\phi(x) = 0
\label{LG}
\end{equation}
where $x=r/\xi_0$, $\tilde \gamma = 2\gamma \Psi_0^2\xi_0^2/\kappa_\phi$ and
$\bar\phi(r) =(\sqrt{\kappa_\phi}/\xi_0)\; \phi(x)$.  This equation is
valid for $x>1$;  for $x<1$, the term $\tilde \gamma/x^2$ is replaced by
$\tilde \gamma$. Manifestly, at large
$x$, $\phi(x) \sim x^{-1/2} \exp(-x(\xi_0/\xi_{\phi}))$ falls with the appropriate
Ornstein-Zernicke form, so
where a non-trivial solution exists, the condensation energy is always finite.  
(As expected, in unscaled units, the vortex halo has a radius $\xi_{\phi}$.)

At the critical point, $\alpha=\alpha_c$
({\it i.e.} $\xi_0/\xi_{\phi}=0$), it is easy to see that  the solution of
Eq. (\ref{LG}) is
\begin{equation}
\phi(x)=\sqrt{1+\tilde\gamma}\ x^{-1} \ \ {\rm for} \ \  x>1.
\end{equation}
(In the screened case, for $x > (\lambda/\xi_0)$, the solution is of the same form,
but with
$\sqrt{1+\tilde\gamma}\to 1$.)  It thus follows that for large $\xi_{\phi}/\xi_0$,
 $\bar \phi$ looks critical for
a large intermediate range, $\xi_{\phi} > r > \xi_0$.  From this, it follows that, as
$\alpha \to \alpha_c^+$,
\begin{eqnarray}
\int d\vec r |\bar \phi|^2 &&\sim 2\pi\kappa_\phi
(1+\tilde\gamma)\log|A\xi_{\phi}/\xi_0|\nonumber\\
\int d\vec r |\bar \phi|^4 &&\sim 2\pi\left(\frac{\kappa_\phi}{\xi_0}\right)^2
(1+\tilde\gamma)^2A^{\prime}
\label{moments}
\end{eqnarray}
where $A$ and $A^{\prime}$ are numbers of order 1.
The first of these results is the essential ingredient
leading to the existence of an avoided critical point in this problem.
Note that the long-range tails of the
vortex profile contribute to, but are not essential to this result;
even if the vortex is screened, $\bar \phi$
at the critical coupling still has a $1/r$ form, so the second moment still diverges
logarithmically as the
critical coupling is approached, although with a somewhat different pre-factor.

For $\alpha$ not so close to $\alpha_c$, we expect a critical value of
$\alpha=\alpha_1^{(0)}$ such that for
$\alpha>\alpha_1^{(0)}$ there are no, non-trivial solutions to the LG equations in the
presence of an isolated
vortex.  It is relatively straightforward to prove that a necessary and sufficient
condition for the existence of
a non-trivial solution is that the quadratic kernel in ${\cal F}$ have at least one
negative eigenvalue, {\it
i.\ e.\/} that
\begin{eqnarray}
\left[-\frac {\partial^2} {\partial x^2} -\frac{1}{x} \frac {\partial} {\partial x}
+\gamma|\bar\Psi(r)|^2
\right]\!\! \phi_0(r) \!\! &=&
\!\!\! \epsilon\phi_0(r)
\end{eqnarray}
has a solution
with $\epsilon<0$.  Thus, $\alpha_1^{(0)}$ is defined implicitly from
the condition $\epsilon=0$.  For $\tilde \gamma=1$ and  the approximate form of the vortex profile introduced
above, 
\begin{equation}
\alpha_1^{(0)}= -0.75\frac{\kappa_\phi}{{\xi_0}^2} \; \; {\rm or} \; \;
\xi_{\phi}= 1.97\;  \xi_0.
\label{alpha1}
\end{equation}
Similar estimates can be obtained for any positive $\tilde \gamma$.

Finally, it is clear from the asymptotic form of the single vortex solutions that
for dilute vortices, the
effects of coupling between vortices are of order
\begin{equation}
J_{inter} \sim \exp[- R/\xi_{\phi}]
\end{equation}
where $R\gg\xi_{\phi}$ is the spacing between vortices.

\section{The Effective action}

To complete the quantum description of the order parameter fluctuations in the
presence of an isolated vortex-line, we define an effective Euclidean action
which includes inter-plane couplings and the simplest possible quantum dynamics.
We will consider the dynamics only of the slow modes of the order parameter $\phi$.
However we will ignore the effects of
the (possibly interesting) coupling to superconducting quasiparticles and other low
energy degrees of freedom - at least in a BCS
$d$-wave superconductor they have a vanishingly small density of states and their
effects on the dynamics of the order parameters of interest
here is likely small. In addition,  since the phase mode of the superconducting order
parameter $\Psi$ decouples at long-wave-length from the rest of the
degrees of freedom, it is reasonable to suppose that quantum fluctuations of the
superconducting order parameter can be
integrated out to produce a small renormalization of the effective parameters;  we thus
consider only static configurations of
$\Psi$, and omit any dependence on $\dot\Psi$. For all the cases of competing order $\phi$
that we are discussing here, the effective
Euclidean action takes the form
\begin{eqnarray}
S[{\phi},{\Psi}]=&&\sum_j\int_0^{\beta} d\tau \int d\vec r
\left\{(M/2) |\dot\phi|^2 \right. \\
&&+{\cal F}[\phi_j,\Psi]
\left. -J_0 (\phi_j^{\dagger}\phi_{j+1} + {\rm h.\ c.\/})\right\}\nonumber
\end{eqnarray}
where $\vec r$ is a 2D vector in each plane, $j$ labels the planes, $\phi_j(\vec r)$
is the order parameter
field in plane $j$, and we only consider the case in which $\Psi_j(\vec r)=\Psi(\vec r)$
is independent of
layer index, {\it i.e.} when we consider a vortex core, we assume the vortex line is
static and precisely
perpendicular to the layers.  Other forms of the $\phi$ dynamics can be considered, but
we believe that, for the most part, the
results will not be qualitatively different.

Our  goal is to obtain explicit expressions for the couplings in the effective
Hamiltonian,
discussed
in Sec. II, above.  In particular, wherever there is a non-trivial solution to the
mean-field equations,
there are clearly a family of equivalent solutions, and important fluctuations which
could potentially restore
the broken symmetry are those which carry the system from one solution to another.

We begin with the case in which the order parameter $\phi$ has  Ising symmetry. In the
Ising case, the classical ground-states are
$\phi=\sigma\bar\phi$, where $\sigma=\pm 1$.  Thus,  at low low temperatures,
the relevant states are of the form $\phi_j(\vec
r)=\sigma_j\bar\phi(r)$.  The effective Hamiltonian has matrix elements,
denoted below by $h$, connecting these classical states with each
other via tunneling processes (``spin flip"). However, while a detailed calculation
of these matrix elements  requires additional analysis to
determine the tunneling paths that locally permit the system to fluctuate from one
ground-state to the other, the end result is quite simple:
the resulting effective Hamiltonian must be of the form of a transverse-field Ising model
\begin{equation}
H_{\rm eff}=h\sum_j\sigma_j^x -J\sum_j\sigma_j^z\sigma_{j+1}^z,
\label{eq:Heff-ising}
\end{equation}
where $J$ is given by
\begin{equation}
J/2J_0 = \int d\vec r |\bar\phi(r)|^2.
\label{J}
\end{equation}
In Eq.\ \ref{eq:Heff-ising},  $h$ is the tunneling
matrix element which therefore depends exponentially on the product of the effective
mass $M^*$ times the barrier
height $E_{\rm cond}$.  Since the effective mass is renormalized by precisely the same
factor,
$M^*/M = J/2J_0$, it follows that
\begin{equation}
\log h \sim -A^{\prime\prime}\sqrt{M^* |E_{\rm cond}|}\sim -\sqrt{\log[\xi_{\phi}/\xi_0]}. \label{eq:h}
\end{equation}
where the prefactor $A''$ is a constant determined by details of the tunneling process.
This implies that this problem approaches the classical 1D Ising-ferromagnet as
$\alpha\to\alpha_c$.

This sort of analysis becomes simpler in
the case in which a continuous symmetry is broken by $\phi$.  For instance, consider
the XY and Heisenberg cases
in which $\phi$ is, respectively, a two or three component vector, while $\bar\phi$,
which is the solution of
the LG equations, can be taken to be a scalar.  Then, the full set of mean-field
solutions can be written as
${\bf \phi}(\vec r)={\bf \Omega}\bar\phi( r)$, where ${\bf\Omega}$ is a unit vector
in order-parameter space.
Now, for the case of weak inter-layer coupling and large ``mass,'' $M^*$, we can
ignore ``amplitude'' fluctuations,
and focus our attention on the soft Goldstone modes by restricting our attention to
field configurations of the
form
\begin{equation}
{\bf\phi}_j(\vec r,\tau)={\bf \Omega}_j(\tau)\bar \phi(r)
\end{equation}
so that the effective Hamiltonian is
\begin{equation}
H_{\rm eff}=\sum_j\frac {|{\bf L}_j|^2} {2M^*} -J\sum_j{\bf\Omega}_j\cdot{\bf \Omega}_{j+1}
\end{equation}
where ${\bf L}_j$ is the angular-momentum conjugate to ${\bf\Omega}$, and $J$ is
given by the same expression, Eq.\  (\ref{J}),
as in the Ising case.
This is the Hamiltonian of a 1D array of quantum rotors, as discussed previously.
Clearly, it follows from Eq.\
(\ref{moments}) that as $\alpha\to\alpha_c$, both $J$ and $M^*$ diverge, making the
system more and more nearly a
classical ferromagnet.

These results flesh out the general physical arguments made in the beginning of this
 paper.  The effect of
amplitude fluctuations have not been included in any of the present
considerations - since ultimately we are
dealing with a 3D quantum system, we believe they are generally less important than the
``phase'' fluctuations we
have explicitly treated.   We have not analyzed the further fluctuation
corrections to test this supposition.

\section{Concluding remarks}

When there is more than one competing order in a system, and the interactions between the
two orders are strong, and a remarkably
large and varied set of phase diagrams are possible\cite{aeppliandme,zhang,vojta}.  The avoided
critical point we
have found here is a particularly striking example.   In the context of
stripe order, the situation is likely to
be even more complex, since several distinct stripe orders have been observed in materials with
stripy tendencies.
Thus, when stripe
order competes with superconductivity,
the full phase diagram should have multiple, possibly nested versions of the relatively
simple phase diagram discussed in the
present paper.  

There are a few additional observations we would like to make.  The existence of vortex
halos has many consequences\cite{gabe}, which we have not
explored, for the character of the thermal
vortex states.  In particular, it gives rise to vortices that effectively have
two\cite{erica} core radii, $\xi_{\phi}$ and $\xi_0$, rather
than a single vortex core radius as in conventional BCS superconductors.
In this context, it is interesting to note that Wang {\em et al}\cite{ong},
from an analysis of the Nernst effect in a variety of high temperature superconductors,
have recently adduced evidence for the existence of well defined vortex excitations at
temperatures well above the superconducting
$T_c$.  Moreover, from an analysis of the magnetic field dependence of the Nernst
coefficient in {\LSCO} and {\YBCO}, they showed that there were two characteristic
magnetic field strengths in the putative vortex liquid state,
which they identified as corresponding to two
distinct vortex core radii.  If this interpretation is accepted, the smaller characteristic field, which they call
$B^*$, is to be associated with the maximal core radius, $R^*\equiv\sqrt{\phi_0/2\pi B^*}$, where $\phi_0=hc/2e$ is
the superconducting flux quantum.  At temperatures well below the zero-field
$T_c$, $B^*$ is typically in the range of 30T, corresponding to $R^*=32\AA$.   Since this length scale is
comparable to the vortex halo radius observed\cite{davis} in STM studies of near-optimally doped {\BSCCO}, it is
very tempting to associate $R^*$ with
$\xi_{\phi}$.  In the Nernst effect experiments, $B^*$ is seen to vary roughly linearly with temperature at low
temperatures, so this identification could be tested by looking for similar temperature dependence of the
vortex halo radius in STM experiments.

Finally, we note that the picture advocated here has also a ``dual"
version\cite{lake,dhl,gabe} in which one or the other
form of stripe order is dominant, and superconducting order is sub-dominant.
In this case, a finite density of topological defects of the stripe state,
\textit{dislocations}, which could
be induced by some form of shear or by disorder, plays the role of the vortex density the
problem considered in the present paper.
Thus, a similar phase diagram as that in Fig. \ref{phasediagram} (with the labels changed)
can be constructed for dislocation induced superconductivity in a stripe ordered
phase.

\section{Acknowledgments}

We thank G.\ Aeppli, J.\ C.\ Davis, A.\ Kapitulnik, S.\ Sachdev, and J.\ T.\ Tranquada  for useful and
stimulating discussions.
This work was supported in part by the National
Science Foundation through the grants No. DMR 01-10329 (SAK, at UCLA), DMR 01-32990
(EF, at the University of  Illinois),   DMR 99 - 71503 (DHL, at UC Berkeley),
and DMR 99-78074 (VO, at Princeton University), and by the David and Lucille
Packard Foundation (VO, at Princeton University).



\begin{thebibliography}{99}
\bibitem{comment-2}
In a crystal with a $C_4$ rotation axis, for instance a tetragonal crystal,
the two nematic axes are related by rotations by $\pi/4$.  For
crystals with a mirror plane symmetry, so long as the preferred stripe direction
is not in the mirror plane, the two nematic axes are related
by reflection through this plane.  In either case, the stripe orientational order
parameter is Ising like.  It is also possible to imagine
more complex stripe orientational order parameters in crystals with a screw symmetry,
which still yield an Ising order parameter;  for
example, in the low temperature tetragonal phase of {\LNSCO}, there are two planes per
unit cell and a four-fold screw symmetry.  It is
possible to imagine more exotic possibilities, such as a crystal with a $C_6$ axis, in
which case the nematic order parameter could be related
to that of a three state Potts model.
\label{ref:comment-2}
\bibitem{comment-3}
In other terms, we are assuming that $B$ is infinitesimally larger than $H_{c1}$,
and that $H_{c1} \to 0$.
\label{ref:comment-3}
\bibitem{critical}  The asymptotic shape of the phase diagram in the neighborhood of
$\alpha = \alpha_1$ derives from the quantum critical
renormalization of $J$ by quantum fluctuations.  For $\alpha<\alpha_1$,  this leads
to a singular dependence of the susceptibility of an
isolated vortex, so that $A^{\prime}\sim [\alpha_1-\alpha]^{z\nu_I}$ where
$z=1$ is the dynamical exponent and $\nu_I=1$ is the correlation length exponent of the
transverse field Ising model.  Consequently, $T_c\sim
B^{1/2}[\alpha_1-\alpha]^{z\nu}$.  Similarly, for $\alpha > \alpha_1$, the susceptibility
of a single vortex does not diverge as $T\to
0$, but rather reaches a large, asymptotic value
$\chi_{\phi}(0) \sim [\alpha-\alpha_1]^{-\gamma_I}$, where $\gamma_I=7/4$.
Consequently,
the zero temperature curve $\alpha_c(B)$ approaches $\alpha_1$ in an extremely singular
fashion (not shown in Fig.\ \ref{phasediagram}):
$\alpha_c(B)-\alpha_1\sim \exp[-A\gamma_I^{-1} \xi_{\phi} B^{-1/2}]$.
\end{thebibliography}
\end{document}